\documentclass{article}
\begin{document}
\title{Field Oscillations in a Micromaser with Injected 
Atomic Coherence}
\author{Mark Hillery and Jozef \v{S}kvar\v{c}ek}
\newcommand{\ket}[1]{{\vert{#1}\rangle}}
\maketitle
\begin{abstract}
The electric field in a lossless, regularly-pumped micromaser 
with injected atomic coherence can undergo a period 2 
oscillation in the steady state.  The field changes its
value after a single atom passes through the micromaser
cavity, but returns to its original value after a second atom
travels through.  We give a simple explanation for this
phenomenon in terms of tangent and cotangent states.  We
also examine the effect of cavity damping on this steady state.
\end{abstract}
\section{Introduction}
%
%
A micromaser consists of a high-Q microwave cavity and a
stream of excited Rydberg atoms which pumps the field in
the cavity.  At any given time, the cavity is either empty
or contains one atom.  Because the field is pumped by only
a single atom at a time, the micromaser is a highly quantum 
mechanical system.  Experiments conducted at the Max Planck
Institute for Quantum Optics in Garching have seen such
quantum phenomena as sub-Poissonian photon statistics and
quantum collapses and revivals~\cite{ex:OneAtomMaser,%
ex:ObservationOfSubPoissonianStatistics}.

Extensive theoretical work on the micromaser has shown
the existence of numerous interesting features.  Among
them are trapping states, which separate the dynamics
into noninteracting blocks of photon 
numbers~\cite{th:Filipowicz}, and
the dependence of steady state of the field on the
injection statistics of the 
atoms~\cite{th:BergouDavidovich}.
The latter occured as part of more general analysis of     
pump noise in lasers and masers, and how to                
minimize it~\cite{th:HaakeTan,th:MarteZoller}.             
Despite all of
this, however, there are still some surprises.
One, found only recently, involves the behavior of the 
electric field inside the cavity.  It has been found by
Briegel, Englert, and Scully~\cite{th:SpectralProperties}, 
and independently by Herzog
and Bergou~\cite{th:Reflection}, that the passage 
of a single atom can cause
the sign of the electric field to flip.  They demonstrated
this by calculating the two-time field-field correlation
function and showing that it can change sign.  For a
micromaser with regular pumping (equal time intervals
between atoms) this effect causes a splitting of the 
spectrum into several equidistant peaks.

In this work the atoms were injected into the cavity
in their upper states.  Another kind of field oscillation
was found by Slosser and Meystre when the atoms are
injected in a coherent superposition of their upper
and lower states~\cite{th:TangentAndCotangentStates}.  
Considering a lossless micromaser,
they found steady states for the field which they called
tangent and cotangent states (the names refer to the 
form of coefficients when the states are expanded in 
the number-state basis), which are confined between
trapping states~\cite{th:Filipowicz}.  
They then went on to examine the
dynamics of the system when the photon numbers were
limited by two trapping states.  They found,numerically, 
that if other trapping states intervened between the 
original two, there were states which one might call steady 
states of period 2; these states are mapped back onto 
themselves not by the interaction with one atom, as a 
normal steady state would be, but by the interaction with 
two atoms.  The electric field of such a state oscillates 
between two values; it changes its value after the passage
of one atom and returns to its original value after the
passage of a second.

Here we would like to give a simple explanation for the
oscillations seen by Slosser and Meystre.  This explanation
hinges on a slight difference in the behavior of tangent 
and cotangent states after the passage of an atom.  In 
addition we wish to see how the oscillatory behavior is
affected by the presence of losses.  In the next section 
we examine the lossless case and in the following section
losses are included.

%
%
\section{Lossless Micromaser}
We shall consider a micromaser in which the pumping atoms
are injected at regular time intervals, with the time
between atoms being $T$.  
Strictly speaking, this assumption is not necessary for our  
analysis in the lossless case to be valid. However, when     
we do consider losses it will be necessary to make           
a particular choice for the injection statistics, and
this is the choice we shall make.
Each atom interacts with the
field for a time, $\tau$, which is much smaller than $T$.
Between atoms the field evolves freely, while during the 
time an atom is in the cavity the interaction of the field
and the atom is governed by the Jaynes-Cummings Hamiltonian
\begin{equation}
H=\omega a^{\dagger}a + \frac{1}{2}\omega (\sigma_{3}+I)
+g(a^{\dagger}\sigma^{-}+a\sigma^{+}),
\end{equation}
where $\omega$ is both the frequency of the cavity mode
and the transition frequency of the atom, and $g$ is the
atom-field coupling constant.  The two-level atom is
described by a two-dimensional state space on which the
Pauli matrixes in the Hamiltonian act.  We denote the
upper state of the atom by $|a\rangle$ and the lower by
$|b\rangle$.  Note that we have assumed that the atom 
and the cavity mode are in resonance.

Let us now suppose that the atom is initially in the state
$\alpha |a\rangle +\beta |b\rangle$, and the field is in 
the state
\begin{equation}
|f\rangle = \sum_{n=0}^{\infty}d_{n}|n\rangle ,
\end{equation}
where $|n\rangle$ is a photon number state with $n$ photons.
If this state evolves for a time $\tau$ under the action
of the Hamiltonian in Eq. (1), then the resulting state
in the interaction picture is
\begin{eqnarray}
|f\rangle (\alpha |a\rangle + \beta |b\rangle ) &\rightarrow &
\sum_{n=0}^{\infty}d_{n}(\alpha c_{n+1}|n\rangle -i\beta s_{n}
|n-1\rangle )|a\rangle \nonumber \\ & & 
+\sum_{n=0}^{\infty} d_{n}(\beta c_{n}|n\rangle -i\alpha s_{n+1}
|n+1\rangle )|b\rangle ,
\end{eqnarray}
where 
\begin{equation}
s_{n}=\sin (g\tau\sqrt{n}) \hspace{2cm} c_{n}=\cos (g\tau
\sqrt{n}) .
\end{equation}
Slosser and Meystre found the tangent and cotangent states by
demanding that the state on the right-hand side of this
equation be a product of the original field state and an
atomic state, i.\ e.\ that
\begin{equation}
|f\rangle (\alpha |a\rangle + \beta |b\rangle ) \rightarrow 
|f\rangle (\alpha^{\prime} |a\rangle + \beta^{\prime} 
|b\rangle ).
\end{equation}
This condition guarantees that the field state is unchanged by
the passage of an atom.  They found that this condition is
satisfied if either $\alpha^{\prime}=-\alpha$, $\beta^{\prime}
=\beta$, and
\begin{equation}
d_{n}=-i\frac{\alpha}{\beta}\cot (g\tau\sqrt{n}/2)d_{n-1},  \label{eq:cot}
\end{equation}
(cotangent state) or if $\alpha^{\prime}=\alpha$, $\beta^{\prime}
=-\beta$, and
\begin{equation}
d_{n}=i\frac{\alpha}{\beta}\tan (g\tau\sqrt{n}/2)d_{n-1} ,  \label{eq:tan}
\end{equation}
(tangent state).  The tangent and cotangent states are
normalizeable only if the sums over number states are
restricted to a finite range.  Expressing both states as
\begin{equation}
|f\rangle = \sum_{n=N_{d}}^{N_{u}}d_{n}|n\rangle ,
\end{equation}
and imposing the conditions $d_{N_{d}-1}=0$ and 
$d_{N_{u}+1}=0$, we find that
\begin{equation}
g\tau\sqrt{N_{u}+1}=p\pi \hspace{2cm} g\tau\sqrt{N_{d}}=q\pi ,
\end{equation}
where $p$ and $q$ are integers.  For tangent states $p$ is even
and $q$ is odd, while for cotangent states $p$ is odd and $q$
is even.

The results of the preceding paragraph imply that for a tangent 
state, $|f_{t}\rangle$,
\begin{equation}
|f_{t}\rangle (\alpha |a\rangle +\beta |b\rangle )\rightarrow
|f_{t}\rangle (\alpha |a\rangle -\beta |b\rangle ),
\end{equation}
and for a cotangent state, $|f_{c}\rangle$,
\begin{equation}
|f_{c}\rangle (\alpha |a\rangle +\beta |b\rangle )\rightarrow
-|f_{c}\rangle (\alpha |a\rangle -\beta |b\rangle ).
\end{equation}
If we have a state which is a superposition of a tangent and a
cotangent state, 
$\xi_{t}|f_{t}\rangle + \xi_{c}|f_{c}\rangle $,
see Figure~\ref{fig:CotTan},
then after one atom passes through the cavity it becomes
$\xi_{t}|f_{t}\rangle - \xi_{c}|f_{c}\rangle $ (where the states 
are defined up to a common sign), and after a second atom it 
returns to the original state.  This state is then periodic with 
period 2 (where time is measured in units of $T$).

How can we see this oscillation of the state?  Because of
the conditions on $N_{d}$ and $N_{u}$, tangent and cotangent
states must exist in nonoverlapping blocks of number states.
This means that the oscillations will not show up in 
observables which commute with the number operator.  On the
other hand, if the blocks are adjacent, then the electric
field operator, which is proportional to $(a^{\dagger}-a)$,
can connect the two blocks, and the effect of the relative
sign flip between them will manifest itself as an oscillation
in the expectation value of the field.  In order to see this
explicitly let us look at the situation considered by
Slosser and Meystre when $g\tau =\pi$, and there is a 
cotangent state at $n=0$ 
(for certain values of the interaction                       
time the vacuum satisfies the conditions necessary to be     
a cotangent state),                                          
a tangent state between 1 and 3,
and a cotangent state between 4 and 8.  The expectation value
of the electric field in this state is 
\begin{equation}
\langle E\rangle = i\sqrt{\frac{\omega}{2V}}\sum_{n=1}^{8}
\sqrt{n}(\rho_{n-1,n}-\rho_{n,n-1}),
\end{equation}
where $V$ is the quantization volume, and $\rho$ is the field
density matrix.  After passage of an atom a relative sign is
introduced between the tangent state and the two cotangent 
states.  This means that the density matrix elements between
different blocks, $\rho_{01}$, $\rho_{34}$, and their complex
conjugates, change sign while the others do not.  This causes
$\langle E\rangle$ to change.  The passage of a second atom
causes these density matrix elements to flip sign again, 
which restores $\langle E\rangle$ to its original value.

The periodicity of the state, then, does have observable
effects.  Besides the field one could also look at an
observables such as $Y_{1}=[a^{2}+(a^{\dagger})^{2}]/2$,
$Y_{2}=i[(a^{\dagger})^{2}-a^{2}]/2$
which appear in the study of some forms of higher-
order squeezing, and which also connect blocks. Because
these connect the number states $|n\rangle$ and $|n+2\rangle$,
rather than $|n\rangle$ and $|n+1\rangle$ as does the
electric field operator, it will lead to a larger number
of density matrix elements which flip sign and can thereby
produce a larger effect.

\section{Micromaser with Losses}
%
%
We now include losses in our system in order to see whether
the steady-state oscillations which we discussed in the
previous section will survive under under these more
realistic conditions.  Because the atom-field interaction 
time $\tau$ is much shorter than
the time the cavity is empty,T, we shall ignore the effect
of field losses during the times atoms are in the cavity.
The decay of the micromaser field 
for the cavity at zero temperature is described
by the master equation
\[ \frac{\mbox{d}\hat\rho }{\mbox{d}t}=-\frac{1}{2}\gamma
\left (a^{\dag}a\hat\rho +\hat\rho a^{\dag}a-
2a\hat\rho a^{\dag}\right )
\]
which has the solution in the number-state representation
%
%
\begin{equation}
\rho_{mn}(t)=\textrm{e}^{-\frac{\gamma t(m+n)}{2}}\sum_{l}
\sqrt{\frac{(m+l)!}{m!}\frac{(n+l)!}{n!}}
\frac{\left (1-\textrm{e}^{-\gamma t}\right )^l}{l!}
\rho_{m+l,n+l}(0).
\end{equation}
%
%
We shall numerically simulate the system by using the
Jaynes-Cummings dynamics to describe the atom-field
interaction and the loss master equation to describe
the field during the periods during which the cavity
is empty.
The values of the micromaser parameters which we
used in our numerical simulations were chosen to
be as close as possible to those which occur in actual 
experiments~\cite{ex:ObservationOfSubPoissonianStatistics}.
The atom-field coupling constant $g$ was set to 
$4.4\times 10^{4}Hz$, the time between two consecutive 
atoms was set to $T=6.666\times 10^{-3}s$,
the cavity loss coefficient $\gamma$ was set to 
$5s^{-1}$, which provided cavity photon storage time 
$T_{\mathit{cav}}=0.2s$, and number of atoms passing 
through the cavity during a single
decay time was taken to be $N_{\mathit{ex}}=30$.
The atom-field interaction time $\tau$ was varied 
in order to provide needed trapping condition, $g\tau =\pi$. 
The values of $\tau$ used in simulations
came very close to the values in actual 
experiment~\cite{ex:ObservationOfSubPoissonianStatistics}.

Two-level atoms in the coherent superposition 
$\alpha \vert a\rangle + \beta |b\rangle$ 
are injected regularly into the micromaser cavity. 
The parameters $\alpha$ and $\beta$ were chosen real
with $\alpha=0.9$. The cavity field was prepared initially
in a state with a cotangent state at $n=0$, a tangent state 
between 1 and 3, and a cotangent state between 4 and 8,
using the same values of $\alpha$, $\beta$ as the pumping
atoms. The starting values for the recurrence
formulas~(\ref{eq:cot},\ref{eq:tan}) were optimized in order
to provide large magnitudes of the observables of interest. 
Then the first atom was injected at $t=0$,
followed by much longer time period $T$ during which the
field decayed. Then the second atom was injected followed
by the cavity decay period and so on. The expectation 
values of observables were calculated after a decay time 
interval just before the injection of the next atom. 

The expectation values of the electric field, operator 
$Y_1$ and $Y_2$ for the case are plotted 
in Figure~\ref{fig:first}.
The interaction time is $\tau =7.14\times 10^{-5}s$ giving 
$\frac{\tau}{T}\cong 10^{-2}$, which justifies the assumption
that we can neglect losses during the atom-field interaction. 
The pumping parameter $\theta_{\mathit{int}}=g\tau
\sqrt{N_{ex}} \cong 17.2$.
The mean value of each operator exhibits period-two 
oscillations, but with decreasing magnitude, because of 
the presence of damping.  These expectation values
eventually reach steady state values which do not
exhibit oscillations. In addition,
we found that the field approaches its steady
state extremely slowly.  This is 
caused by the presence of the trapping states which
separate the total Fock space for the Jaynes-Cummings
evolution into independent blocks.
The probability flow between the subspaces in which the
cotangent and tangent states are located occurs only because
of the loss process, and it is very small.

We now look at the density matrix itself.
The photon number distribution of the field, given by the
diagonal density matrix elements, after the 
passage of different numbers of atoms is shown in 
Figure~\ref{fig:second}.  Note that the presence of 
damping drives the field toward lower photon numbers.
In Figure~\ref{fig:third} are plotted absolute values of the
cavity field density matrix elements. Initially, because
the system starts in a pure state, the density
matrix has off-diagonal peaks. Because of the loss process, 
the off-diagonal elements of the density matrix decay
as it approaches its new steady state.
This new steady state is very different from the initial
cotangent-tangent state, and this explains 
the deterioration of the oscillations in the expectation
values of the operators, because they depend on the 
presence of the tangent and cotangent states which 
exhibit a relative phase oscillation.

Our conclusions on the effects of damping are consistent with
the results of Slosser, Meystre, and Wright~\cite{th:SlosserMeystre}.  
They considered
a micromaser with Poissonian pumping and found that 
tangent and cotangent states maintain their integrity only
in the limit of very large $N_{ex}$, i.\ e.\ very weak damping.
For the damping which occurs in experiments, the damping 
drastically alters the character of the steady-state field.

\section{Conclusion}
%
%

We have shown that the period-two oscillations which exist in
a lossless micromaser pumped by atoms in a coherent
superposition of states are due to a relative sign
change undergone by tangent and cotangent states upon
passage of an atom through the cavity.  These oscillations
can only occur in observables, such as the electric field,
which do not commute with the photon number operator.  The
addition of damping changes the oscillations from a 
steady-state to a transient phenomenon, but one with a rather
long lifetime.
\vskip 5truemm
\parindent 0truemm
Contact: Department of Physics and Astronomy, Hunter College
of the City University of New York, 695 Park Avenue, New York,
NY 10021, U.S.A.
\bibliographystyle{prsty}
%
%

%
%
\begin{figure}[p]
\vskip 1truecm
\caption{Cotangent and tangent states. The state $\ket{k}$
        is $\pi$ trapping state, $g\tau\sqrt{k}=\pi$.}   
                                                        \label{fig:CotTan}
\end{figure}
%
%
%
\begin{figure}[p]
\vskip 1truecm
\caption[operators]{(a): Evolution of the expectation value of 
        electric field $\langle E\rangle$ with respect to the
        number of atoms $n$ which passed through the cavity,
        (b) shows the evolution of $\langle Y_1\rangle$ and (c)
        shows the evolution of $\langle Y_2\rangle$. 
        }                                                   \label{fig:first}
\end{figure}
%
%
\begin{figure}[p]
\vskip 1truecm
\caption[distribution]{Photon number distribution 
        $P_n =\rho_{nn}$ of the cavity field for the
        initial state (a), after interaction with 20 atoms (b),
        and after interaction with 100 atoms (c).
        }                                                  \label{fig:second}
\end{figure}
%
%
\begin{figure}[p]
\vskip 1truecm
\caption[matrices]{Moduli of cavity field density matrix, 
        (a) shows 
        initial state, the nondiagonal peaks are apparent.
        (b) shows the state after interaction with 10 atoms,
        (c) after 20 atoms, (d) after 100 atoms.
        }                                                   \label{fig:third}
\end{figure}

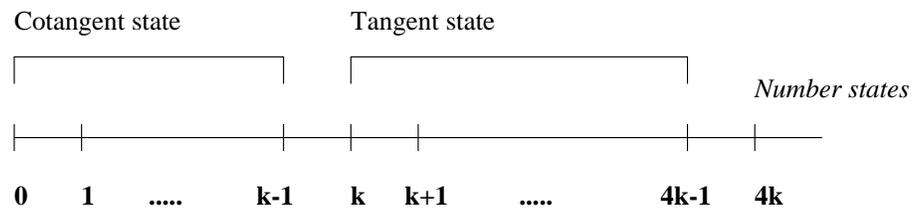
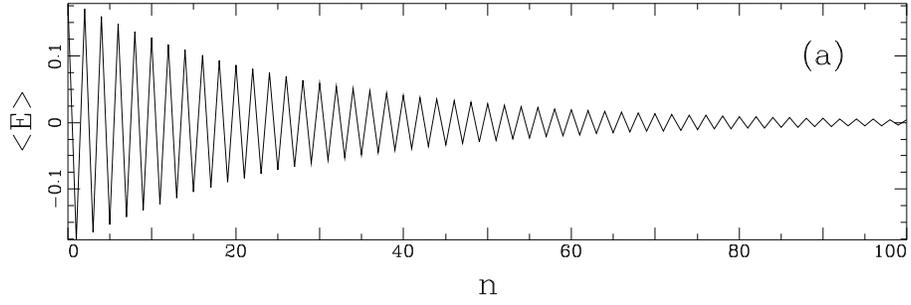
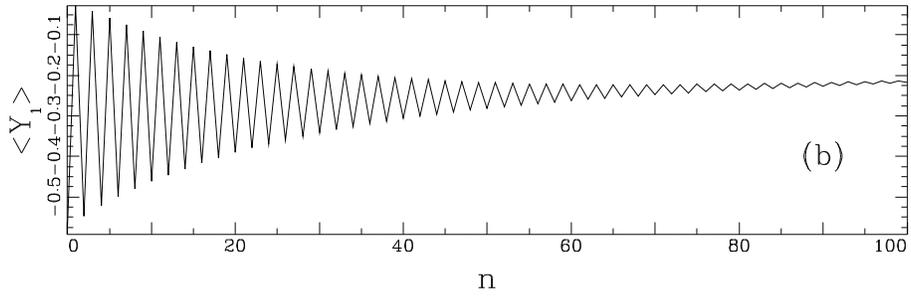
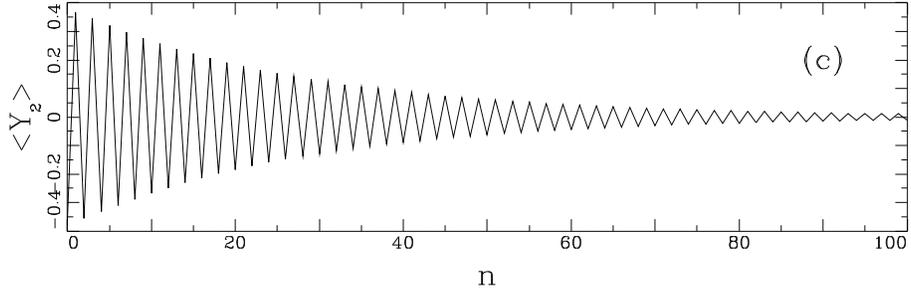
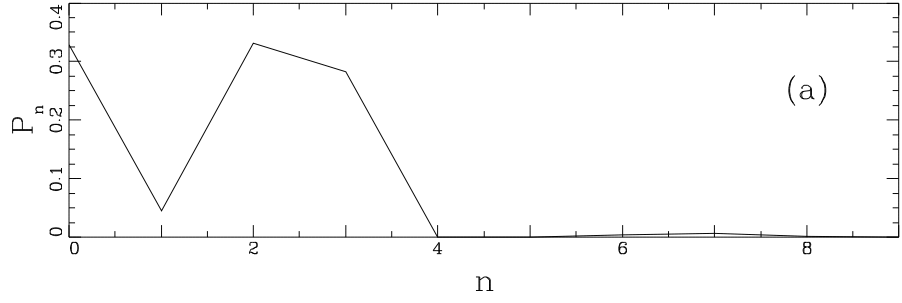
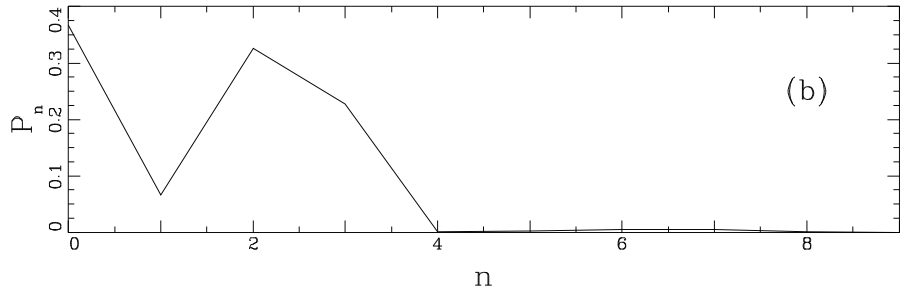
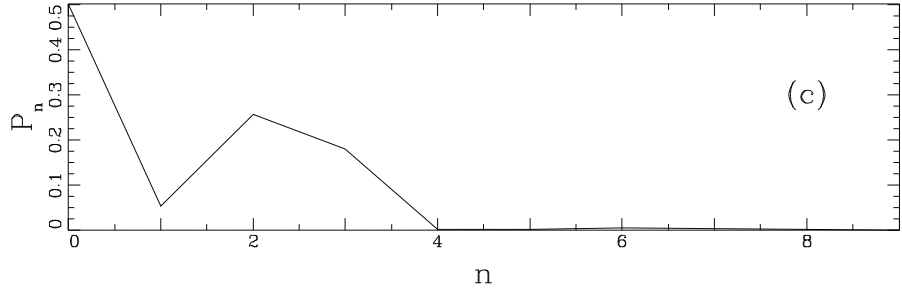
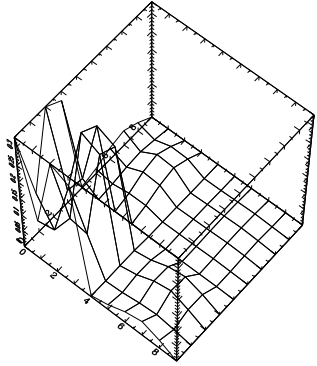
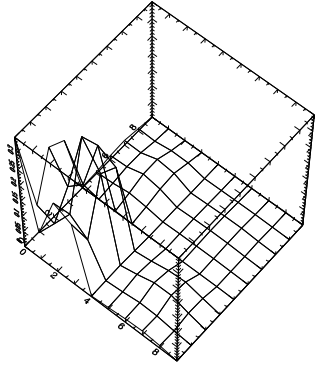
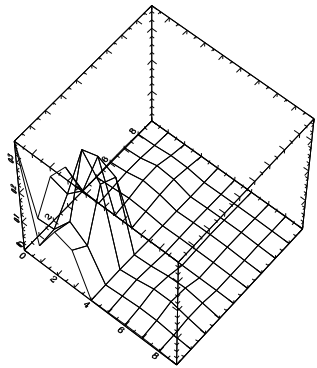
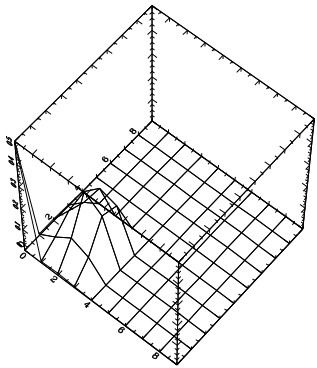
\end{document}